\documentclass[a4paper,11pt]{article}
\usepackage{aaskaiid}
\usepackage{orcidlink}
\usepackage{xeCJK}

\title{Magnetic Reconnection in Galaxy Clusters}
\ShortTitle{Magnetic Reconnection in Galaxy Clusters}
\ShortName{Akahori et al.}
\author[1,2]{Takuya Akahori (赤堀~卓也) \orcidlink{0000-0001-9399-5331}}
\emailAdd{takuya.akahori@nao.ac.jp}
\author[1,3]{Kohei Kurahara(藏原~昂平) \orcidlink{0000-0003-2955-1239}}
\emailAdd{kurahara.kohei.i7@f.mail.nagoya-u.ac.jp}
\author[4]{Shin-ya Nitta (新田~伸也) \orcidlink{0000-0001-9838-7765}}
\emailAdd{snitta@a.tsukuba-tech.ac.jp}
\author[5]{Haruka Sakemi \orcidlink{0000-0002-4037-1346}}
\emailAdd{sakemi@yamaguchi-u.ac.jp}
\author[6]{Hiroki Akamatsu}
\emailAdd{hirokia@post.kek.jp}
\author[7]{Mami Machida \orcidlink{0000-0001-6353-7639}}
\emailAdd{mami.machida@nao.ac.jp}
\author[8]{Motokazu Takizawa}
\emailAdd{takizawa@sci.kj.yamagata-u.ac.jp}
\affiliation[1]{Mizusawa VLBI Observatory, National Astronomical Observatory of Japan, 2-21-1 Mitaka, Tokyo 181-8588, Japan}
\affiliation[2]{SOKENDAI, Shonan Village, Hayama-machi, Miura-gun, Kanagawa 240-0193 Japan}
\affiliation[3]{Kobayashi-Maskawa Institute for the Origin of Particles and the Universe (KMI), Nagoya University, Furo-cho, Chikusa-ku, Nagoya, Aichi 464-8601, Japan}
\affiliation[4]{Tsukuba University of Technology, Japan}
\affiliation[5]{Graduate School of Sciences and Technology for Innovation, Yamaguchi University, Yoshida 1677-1, Yamaguchi 753-8512, Japan}
\affiliation[6]{QUP, KEK, 1-1 Oho, Tsukuba, Ibaraki 300-3256, Japan}
\affiliation[7]{Division of Science, National Astronomical Observatory of Japan, 2-21-1 Mitaka, Tokyo 181-8588, Japan}
\affiliation[8]{Department of Physics, Faculty of Science, Yamagata University, Kojirakawa-machi 1-4-12, Yamagata 990-8560, Japan}

\abstract{
Galaxy clusters contain an intra-cluster medium (ICM) with temperatures of tens of millions of Kelvin. Cosmological structure formation simulations show that this diffuse gas is heated not only by adiabatic gravitational compression but also by shock waves and turbulence generated during mergers of galaxy groups and clusters. These processes are expected to produce magnetic fields and cosmic rays, observed through synchrotron polarization. One structure formed during cluster evolution is the cold front, a contact discontinuity created when colder gas moves transonically through hotter gas. Using MeerKAT, GMRT, and ATCA, we recently discovered radio emission along cold fronts in two galaxy clusters, with spectra indicating re-acceleration at the discontinuity. This presents a new puzzle because the standard mechanism in galaxy clusters, Fermi acceleration, is not naturally expected there. We propose magnetic reconnection as the re-acceleration mechanism. Compression and stretching of magnetized plasma at the discontinuity can generate current sheets that trigger reconnection, as also suggested by simulations. With AA*, we will probe broadband radio spectra at high spatial resolution to constrain where re-acceleration occurs. Polarization measurements will reveal magnetic-field structures and clarify the conditions required for magnetic reconnection.
}

\begin{document}
\maketitle

\section{Introduction}
Galaxy cluster is the largest astronomical object in the Universe. It contain the intra-cluster medium (ICM) with temperature of tens of millions of Kelvin. Such high-temperature gas is important evidence for the existence of dark matter, which governs the gravitational potential. However, understanding how the gas has been heated remains a fundamental question in the structure formation of the Universe. Cosmological simulations have shown that galaxy clusters form through the mergers of smaller systems such as galaxy groups and clusters at the nodes of the cosmic web. The ICM has heated by not only adiabatic compression due to gravity but also shock waves and turbulence generated in the mergers. 

One of the interesting structures formed in the structure formation is called a cold front, which is recognized relatively recently with high-resolution X-ray imaging. The cold front is a contact discontinuity created when, for example, colder ICM is in transonic motion through hotter ICM. The process known as sloshing is one of the representative mechanisms responsible for the formation of cold front. Figure 1 introduces an example of numerical simulations \citep{2010ApJ...717..908Z}. During the process in which a sub-cluster with lower-temperature ICM collides and merges, the cooler ICM falls into the hotter ICM of the host cluster. In addition, if there is radiatively-cooled ICM around the center, it oscillates due to a shake induced by the collision, forming cold fronts.

\begin{figure}[tbp]
    \centering
	\includegraphics[width=\linewidth]{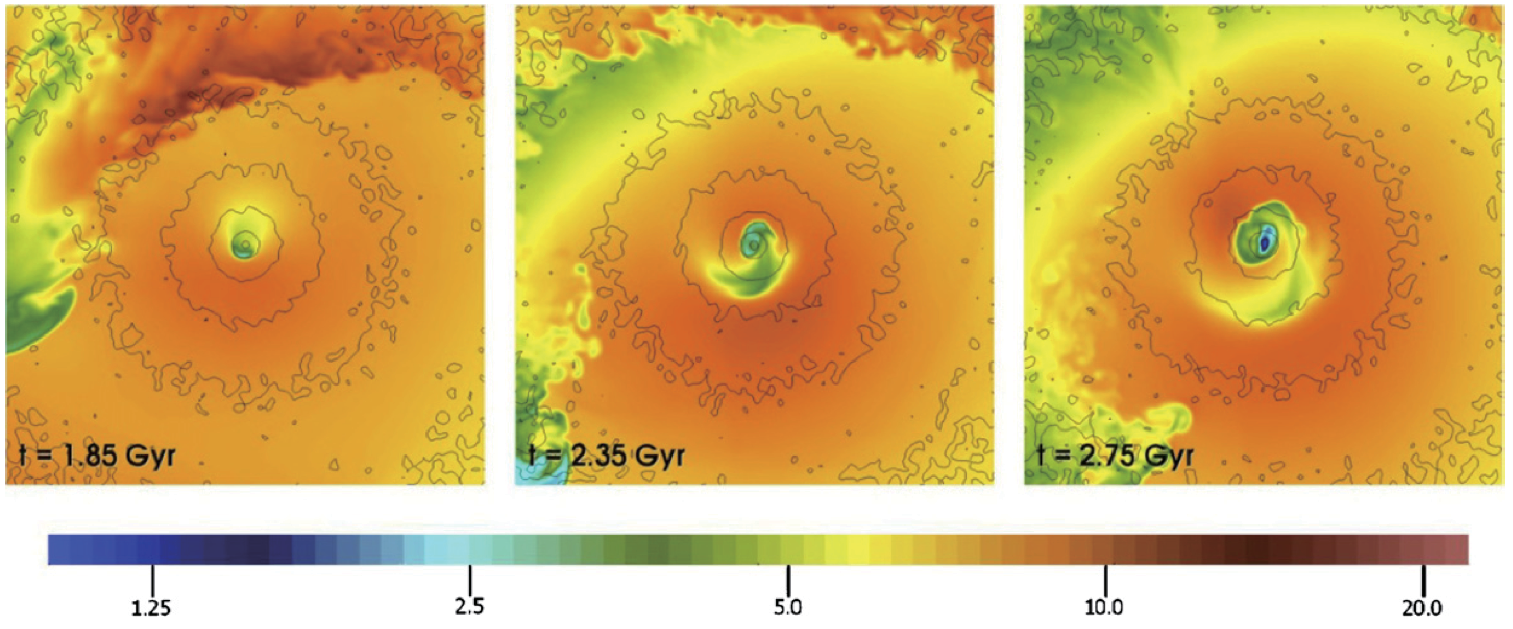}
    \caption{Simulated gas sloshing during a cluster merger \citep[a case R20b1000gc; see][ for details]{2010ApJ...717..908Z}. Color indicates gas temperature in keV with dark matter density contours overlaid. Each panel is 1 Mpc on a side.
    }
    \label{fig:Sloshing}
\end{figure}

The shock waves and turbulence are expected to produce magnetic fields and cosmic-rays, which has been observed as synchrotron polarization \citep[See e.g.,][for reviews]{2012A&ARv..20...54F, 2018PASJ...70R...2A, 2019SSRv..215...16V}. Diffuse radio emissions, such as radio halos and relics in galaxy clusters, are thought to originate from cosmic-ray electrons accelerated in situ by shock waves and turbulence via first- and second-order Fermi process, respectively \citep[See][for collections in this science book]{Vernstrom01.2026.SKA, Vacca01.2026.SKA, AritraBasu01.2026.SKA, Kurahara01.2026.SKA}. It is well known that the spectral index steepens as the electron population ages \citep[See e.g.,][references are therein]{2021Natur.593...47C, 2023PASJ...75S.138K, 2024PASJ...76L...8K, 2026PASJ...78..137K, 2025ApJ...992...14S}.

\section{Recent Observations}

Using MeerKAT, GMRT, and ATCA, we found radio emissions along the cold fronts of two galaxy clusters. The radio spectra indicate cosmic-ray (re-)acceleration rather than simple aging. Below, we briefly summarize the observational results.

\subsection{MRC B0600-399 / Abell 3376}

\begin{figure}[btp]
    \centering
	\includegraphics[width=\linewidth]{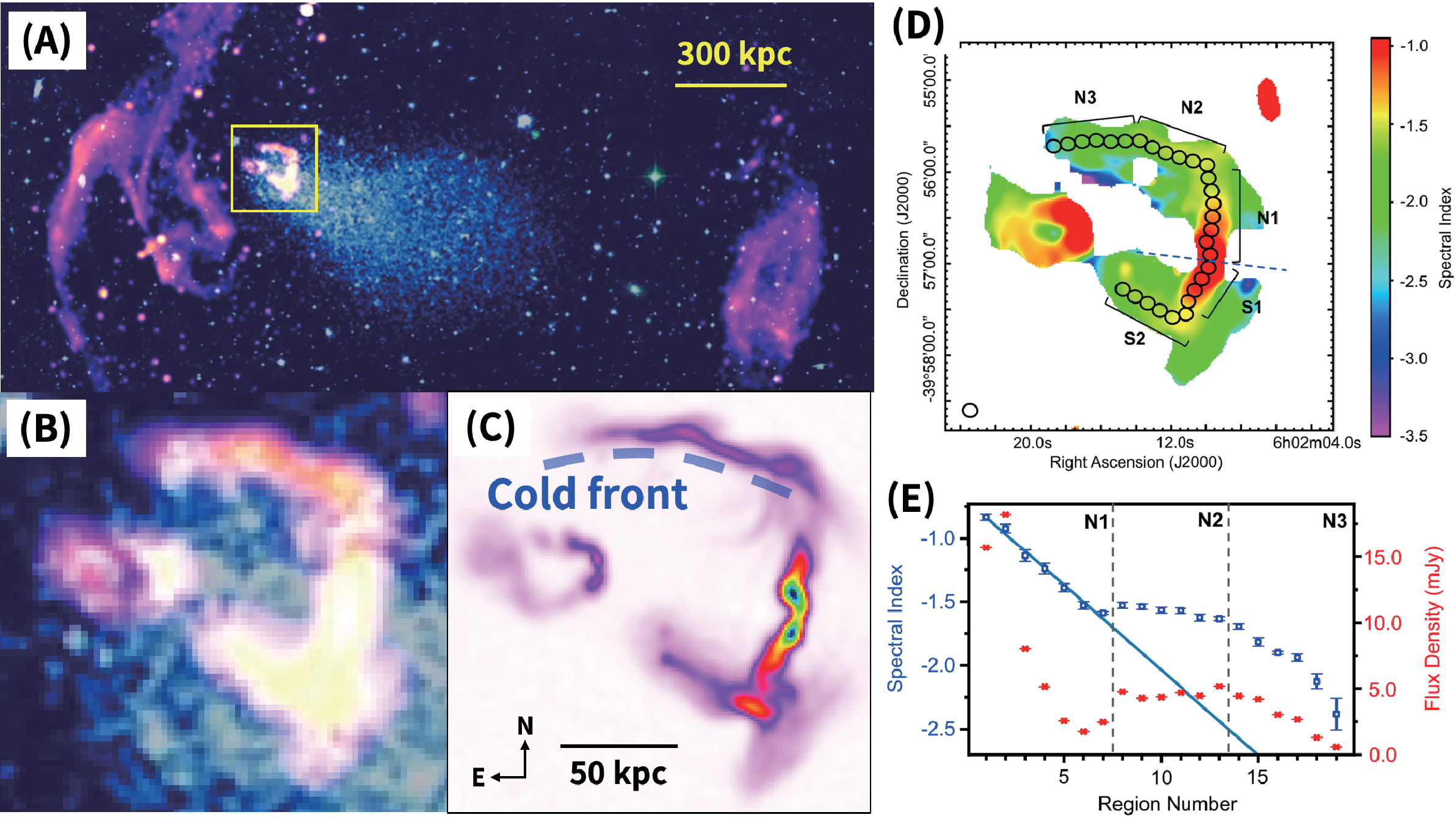}
    \caption{(A) An overview of Abell 3376. The cyan and magenta show the X-ray (XMM-Newton) and radio (MeerKAT 1.28 GHz) surface brightness maps, respectively. (B) A zoom-in view around MRC 0600-399 (yellow box in (A)). (C) An extended radio emission of MRC 0600-399, where the expected position of the cold front is shown as the dashed line. The image was taken with MeerKAT at 1.28 GHz with the beam size $5.80 \times 5.48$ arcsec$^2$ and the image rms 4.2 $\mu$Jy/beam. (D) The radio spectral index map. (E) The radial profile of the index (blue, left scale) and the flux density (red, right scale) along the jet. All figures from \citet{2021Natur.593...47C, 2023PASJ...75S..97C}
    }
    \label{fig:B0600-399}
\end{figure}

Abell 3376 is a nearby ($z = 0.0461$) galaxy cluster. It has an elongated X-ray morphology and a pair of Mpc-scale radio relics, suggesting that Abell 3376 is in a late stage of minor merger. At the eastern rim of the X-ray surface brightness, a jump between the cold-dense gas and the hot-thin gas, i.e. the cold front, is identified. The blurred interface of the cold front is explained by the projection effect. Numerical simulations of cluster merger suggest that the overall X-ray morphology can be explained by a head-on merger with the mass ratio around 1:6 and that the inclination angle of the merger axis with respect to the sky plane would be a few to several tens of degrees \citep{2013MNRAS.430.3249M}.

At the center of the smaller subcluster, there is a radio galaxy MRC 0600-399, which is the second brightest cluster galaxy (BCG) of Abell 3376. This giant elliptical galaxy is ejecting radio jets. Recently, we observed MRC 0600-399 using MeerKAT at 1.28 GHz, and found a 90-degree bend of the jets at the cold front \citep{2021Natur.593...47C}. Figure 2 shows the total intensity map of MRC 0600-399, where N1 extends from the 2nd BCG and N2 and N3 are expected to be along the cold front. In the N1 and N3 regions, the radio spectral index is steepening along the jet. Such a steepening is ascribed to the cosmic-ray aging due to cooling. Meanwhile, we found that the index profile has a plateau at N2 region, suggesting that cosmic-ray (re-)acceleration is taking place. 

Based on the global picture of merger described above, the subcluster containing MRC 0600-399 appears to be passing through the main cluster from west to east. Thus, we expect that the ICM flows from east to west at the outer region of the cold front. However, strangely, both the northern and southern jets direct to east, opposite from the expected flow. Such reverse-flow tails of AGN jets at the cold front has never been reported in the literature, suggesting the existence of the physical mechanisms not revealed yet. We performed MHD simulations of jet propagation under the scenario that the jet is being deflected by ordered magnetic fields along the cold front. We found that a coherent magnetic field of the order of 10 µG along the cold front gives enough magnetic pressure and tension and is sufficient to influence the jet dynamics, reproducing the jet bending. In the scenario, the jet extends along magnetic field lines that are aligned with the cold front, which naturally explains why the jet remains narrowly collimated even after bending. Moreover, the simulation successfully reproduces the radio brightness of the jets, which are ejected both upward and downward, weakens partway along their length but becomes stronger again at the bending point. These physical interpretations are generally consistent with estimates derived from observations. However, significant uncertainties remain, as the results depend on the poorly constrained population of cosmic-ray electrons and on the details of the energy partition processes. See \citet{2021Natur.593...47C} for detailed discussion.

Our latest study of polarization using three-dimensional Faraday RM synthesis \citep{2025ApJ...992...14S} unveiled that N2 and N3 regions does not continuously connect in the Faraday depth domain. Since this region does not display significant variation in thermal electron density, the gap is most likely caused by a change of the line-of-sight magnetic field component. The total intensity map shows no major disruption of the jet structure in this area. It means that the observed Faraday depth changes trace the morphology of the large-scale magnetic field along the cold front, rather than within the jet itself. It is possible that cosmic-ray electrons accelerated by the jet transition into the cold front magnetic field via mechanisms such as perpendicular diffusion or magnetic reconnection, producing synchrotron emission. See discussion of \cite{2025ApJ...992...14S} for details.

\subsection{MRC B1407-425 / CIZA J1410.4-4246}

\begin{figure}[tbp]
    \centering
	\includegraphics[width=\linewidth]{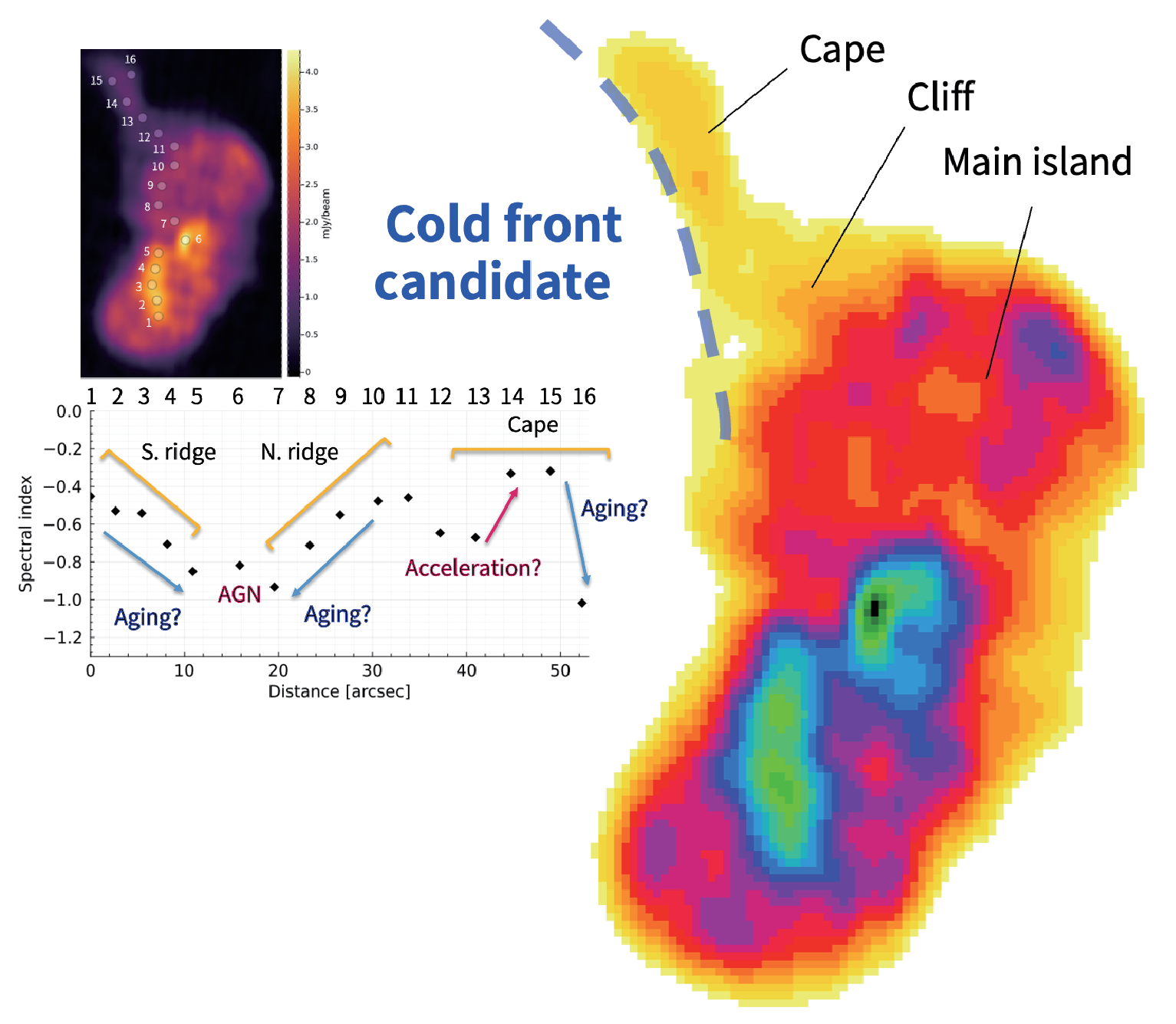}
    \caption{An extended radio emission of MRC B1407-425. The image was taken with ATCA at 5.5 GHz with the beam size $3.40 \times 1.95$ arcsec and the image rms 7.4 $\mu$Jy/beam; Akahori et al. in preparation).}
    \label{fig:B1407-425}
\end{figure}

CIZA J1410.4-4246 is a nearby ($z=0.049$) medium-massive ($4 \times 10^{14}$ ${\rm M}_\odot$) cluster. Overall X-ray shape is regular except for some minor substructures near the center. Previous radio surveys (TGSS, GLEAM, SUMSS) reported only the central unresolved source (MRC B1407-425), and such a featureless appearance would be a reason of attracting less attention from researchers in the past. Recently, this cluster was observed in the MeerKAT galaxy cluster legacy survey (MGCLS). It shows that the central radio source is an arc-shaped diffuse source, yet detailed studies have never been made in the literature.

We conducted multi-frequency observations of MRC B1407-425, at 750 MHz with GMRT, and 2.1 GHz, 5.5 GHz, and 9.0 GHz with ATCA. Figure 3 shows a high resolution ($3.''40 \times 1.''95$) image of the total intensity of MRC B1407-425. It revealed that MRC B1407-425 consists of the central diffuse emission (the main island) and an elongated substructure (the cape) connecting with a brightness gap at the root of the cape (the cliff). The radial profile of the spectral index has a plateau around the cliff and suddenly flattens at the cape, suggesting cosmic-ray (re-)acceleration. Interestingly, the cape has the high fractional polarization up to 70\% with the polarization angle that suggests the parallel magnetic field to the major axis of the cape (not shown).

We checked the X-ray surface brightness taken with XMM-Newton, and realized that there is a sloshing substructure around the center. Since there is no strong support of the existence of temperature jump at the interface between the sloshing gas and the surrounding gas, it is likely that the interface is a cold front. Surprisingly, the cape aligns well with the interface. If the interface is a cold front, similar to MRC B0600-399 in Abell 3376, MRC B1407-425 in CIZA J1410.4-4246 also exhibits the diffuse radio emission along the cold front, where traditional shock acceleration cannot be applied to explain cosmic-ray (re-)acceleration suggested from the spectral index anti-aging. We need an acceleration mechanism at the cold front.

\section{Theoretical Prospect}

This is a new mystery because the standard acceleration mechanism in galaxy clusters, i.e. Fermi acceleration, is not simply expected at the cold front. Here we propose magnetic reconnection as the re-acceleration mechanism. Magnetic reconnection has been studied intensively on the solar surface and wind, but has rarely been discussed in the environment of galaxy clusters. Magnetic reconnection is the process of rearranging magnetic-field lines.

\subsection{Properties of ICM as the boundary conditions of magnetic reconnection}

We first recap an importance of magnetic field for the dynamics of the ICM. For this purpose, the Alfven velocity is a useful measure. For the cluster environment, the Alfven velocity is scaled to be 
\begin{equation}
v_{\rm A} \sim 69~{\rm km/s} \left( \frac{B}{\rm \mu G} \right) \left( \frac{n}{\rm 10^{-3}~cm^{-3}} \right)^{1/2},
\end{equation}
which is comparable to or slower than the ICM bulk velocity and turbulence velocity. Thus, the flow is in the super Alfvenic motion, i.e. magnetic field is not dynamically important. However, the reaction of magnetic field is important when we consider a scale smaller than the Alfven length, where the Alfven time can be shorter than the eddy turnover time and the energy is effectively transferred by Alfven wave rather than eddy cascade. In other words, magnetic field is dynamically important when we consider the timescale sufficiently larger that the Alfven time
\begin{equation}
t_{\rm A} \sim l/v_{\rm A} \sim 2\times 10^7~{\rm years}~\left( \frac{B}{\rm \mu G} \right)^{-1} \left( \frac{l}{\rm 10~kpc} \right) \left( \frac{n}{\rm 10^{-3}~cm^{-3}} \right)^{1/2}.
\end{equation}

Under the existence of magnetic fields, electrons and ions of a plasma are gyrating around the field lines and they cannot easily cross them. It results in the magnetic field effectively frozen into the plasma. Such a frozen plasma follows the ideal MHD equations, where the electrical conductivity $\sigma$ is large ($\infty$). High temperature plasma is known to possess a large $\sigma$ value, i.e., the magnetic diffusivity is close to zero, so that the ICM is believed to satisfy the ideal MHD conditions. This means that the ICM has a high magnetic Raynolds number, $R_{\rm M} \sim 10^{23-29}$ (1 $\mu$G, 100 km/s, 1 kpc) and behaves like adiabatic gas with a large Prandtl number $P_{\rm r} = R_{\rm M}/R_{\rm e} \sim 10^{20}$, where $R_{\rm e}$ is the Reynolds number. Then, it has been proven that magnetic reconnection never happens under the ideal MHD \citep{1958AnPhy...3..347N}, so that magnetic reconnection is not expected in the ICM. 

However, violent plasma motions may locally violate the frozen-in condition, giving rise to magnetic diffusion and a transition to the resistive MHD regime, where magnetic reconnection could occur. The non-linear evolution of magnetic reconnection depends on the magnetic diffusivity $\eta$, and the condition is often described with the Lundquist number, $S=Lv_{\rm A}/\eta$, where $L$ is the spatial scale. The ICM is expected to have a large Lundquist number. Classically (1) the Sweet-Parker type reconnection takes place with a relatively low Lundquist number and it forms plasmoids if $\eta$ is spatially uniform, or (2) the Peschek type reconnection takes place and the reconnection jets accompany slow shocks at both sides of the jets if $\eta$ is spatially localized. A known longstanding issue is the long timescale of these processes. The conventional Sweet-Parker (SP) model is too slow to be considered viable in this context, where the timescale of the SP model is $t_{\rm SP} \sim (R_{\rm M})^{1/2} t_{\rm A}$. Even the Petschek model is not fast enough, $t_{\rm P} \sim 8 \log (R_{\rm M}) t_{\rm A}/\pi \sim 100~t_{\rm A}$. Moreover, this refers to the timescale of MHD energy release, not to the timescale of particle acceleration. 

The currently accepted paradigm for fast reconnection in high Lundquist number plasmas is the plasmoid-mediated reconnection regime, which occurs when $S \gtrsim$ a few $\times 10^4$ in a MHD system. In this regime, long or thin current sheets become unstable to the plasmoid instability, producing multiple reconnection sites and enabling fast energy release. In addition, a new reconnection model that can cover a wide range of Lundquist number has been proposed \citep{2001ApJ...550.1119N, 2007ApJ...663..610N}. Reconnection that begins locally in a long current sheet expands explosively and self-similarly at fast-mode speed of the surrounding circumstance, and finally it forms a huge energy conversion structure. Depending on the Lundquist number, a fast (high power) structure with a long slow-mode shock like the Petschek model (for low Lundquist number) and a slow (low power) Sweet-Parker-like structure with multiple reconnection points (for high Lundquist number) are predicted.

\subsection{Numerical Simulations}

\begin{figure}[tbp]
    \centering
	\includegraphics[width=0.3\linewidth]{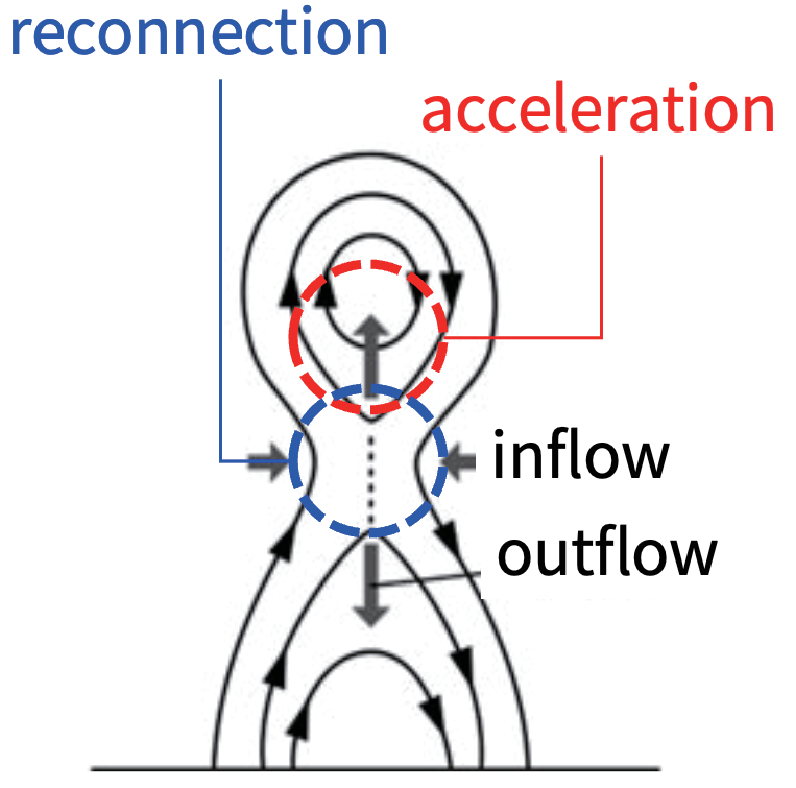}
	\includegraphics[width=0.65\linewidth]{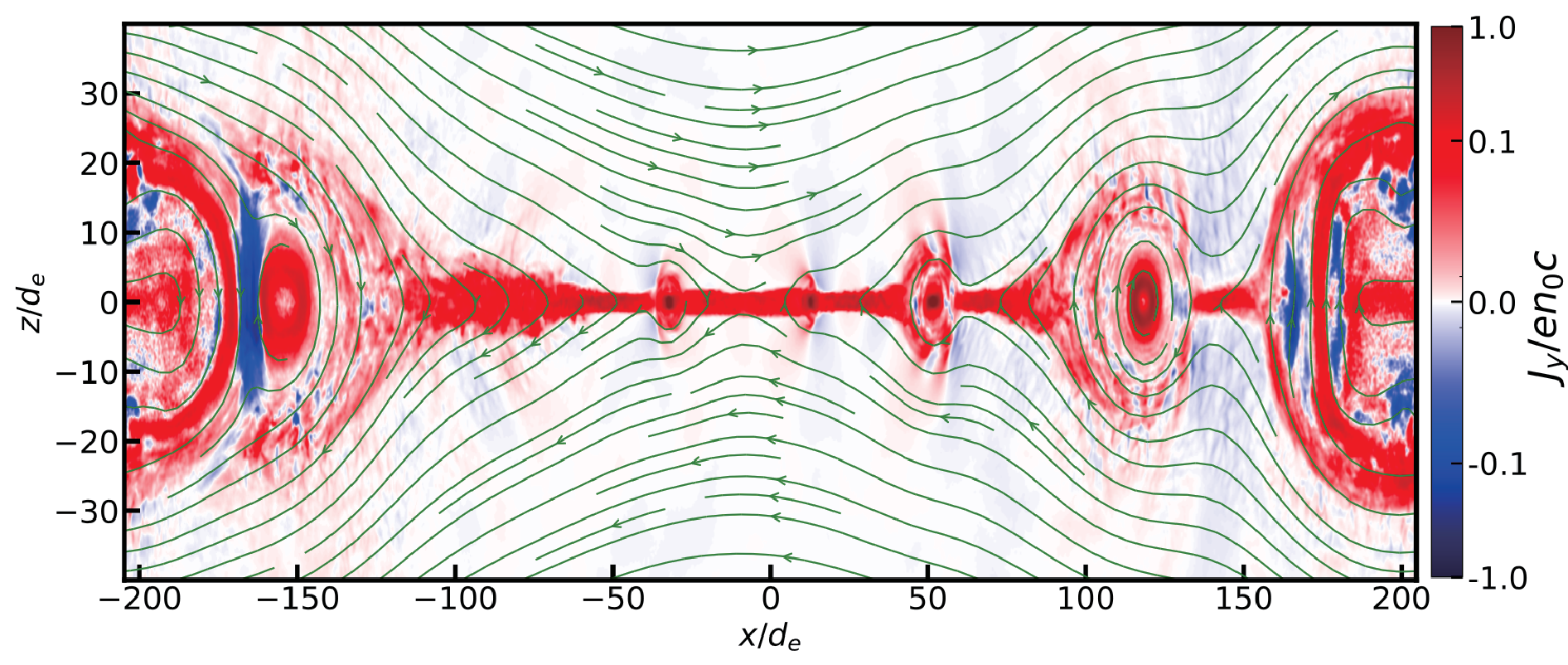}
    \caption{Left: a scketch of the earing mode instability at the Solar surface. Right: the PIC simulation of magnetic reconnection for the turbulent ICM situation.}
    \label{fig:PIC}
\end{figure}

\begin{figure}[tbp]
    \centering
	\includegraphics[width=\linewidth]{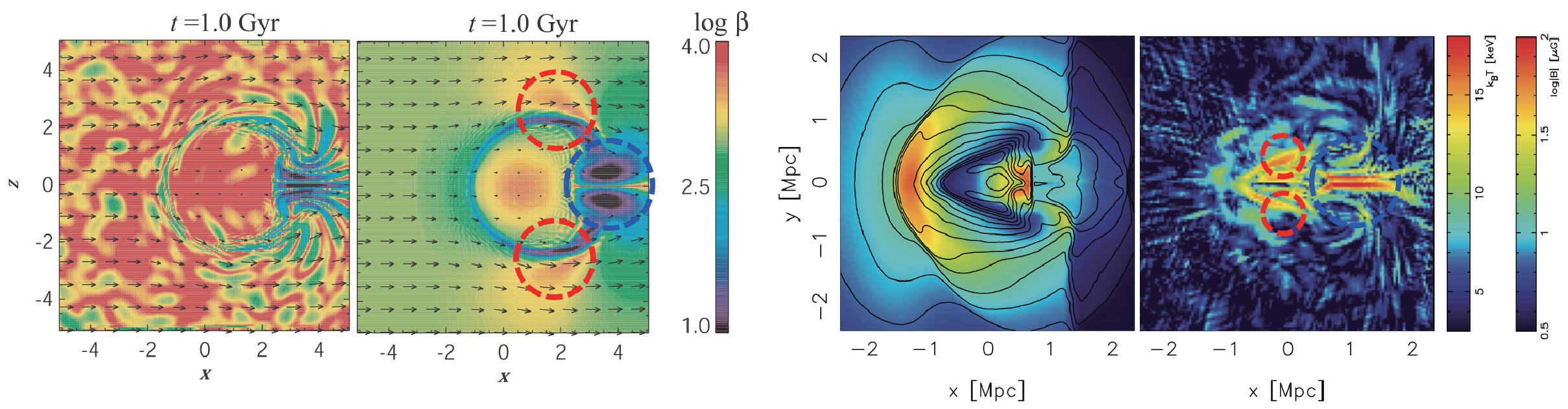}
    \caption{MHD simulations of merging galaxy clusters. Left: The plasma $\beta$ for turbulent and coherent initial magnetic fields \citep{2007ApJ...663..816A}. Right: ICM temperature and field strength  \citep{2008ApJ...687..951T}. The red dashed circles indicates the regions with low plasma $\beta$ along the cold front. The blue dashed circles are also a candidate of the reconnection site.
    }
    \label{fig:MHD}
\end{figure}

Recently, Particle-In-Cell (PIC) simulations investigated the realization of the tearing mode (plasmoid) instability and plasmoid formation in the cluster environment \citep{2025ApJ...979L..15G} (Fig. \ref{fig:PIC}). The plasmoid instability is expected as faster energy release with a higher reconnection rate and more rapid reconnection. They claimed the realization of magnetic reconnection and efficient particle acceleration in collisionless and nonrelativistic plasmas on a realistic timescale. 

A key condition in the above simulations is the boundary pressure balance between magnetic and thermal forces, i.e. the plasma $\beta$ value is close to unity. Meanwhile, in a global scale, it is known that the ICM typically possesses $\beta \sim 100$ for $O(1)$ $\mu$G. This means that it is less likely that magnetic reconnection caused by the plasmoid instability occurs on a global scale. On the other hand, MHD turbulence simulations have shown that even when the global plasma $\beta$ is $O(100)$, it is not surprising that disturbances including gas compression can locally amplify the magnetic field strength by a factor of ten, creating discrete regions where the plasma $\beta$ approaches unity. In the future, it is important to quantitatively examine whether the total energy released by locally occurring reconnection events is consistent with the luminosity inferred from radio observations and with the energy of cosmic rays.

Another notable simulations are MHD simulations of merging clusters. It is interesting that the MHD simulations predicted strong magnetic fields of O(10) $\mu$G, or the plasma $\beta \sim 1$, at the cold front located at the side of the sub-cluster core 
\citep{2007ApJ...663..816A,2008ApJ...687..951T} (Fig. \ref{fig:MHD}). These strong magnetic fields are generated by stretching of the field lines as well as the compression of the gas. Therefore, the condition favorable for magnetic reconnection is met at the cold front. In the future, to reveal where the low $\beta$ regions emerge and to quantitatively estimate the plasma $\beta$, it is ideal if we conduct MHD simulations of the gas sloshing for the models of MRC B0600-399 and MRC B1407-425.

\section{Study of Magnetic Reconnection with SKA}
 
If magnetic reconnection occurs at the cluster scale, this would be a new paradigm for the non-thermal aspects of galaxy clusters. Potential observation targets are the cold fronts at which particle (re-)acceleration is expected from the spectral anti-aging seen in diffuse radio emission. Obviously, MRC B0600-399 and MRC B1407-425 are interesting ones.

To study the observational feasibility, hereafter we consider a source at the declination of $-40$~degree. Such a source can be seen with the elevation above 45~degree for at least 8 hours. Based on the SKAO's sensitivity calculator with those values, at Band 2 1.31 GHz, the synthesized beam size and the sensitivity are $0.''75 \times 0.''22$ and 6.86 $\mu$Jy/beam, respectively, for AA*. Such a specification can deliver a similar image noise level seen in Figs. \ref{fig:B0600-399} and \ref{fig:B1407-425}, but with a factor of several, dramatically-improved resolution. This resolution allows us to see a typical Alfven scale, where magnetic field becomes important, 100 pc or less, corresponding to $0.''5$ for the distance of 43 Mpc ($z=0.01$). Therefore, we can probe the MHD regime for the first time. Even for AA4, the synthesized beam size and the sensitivity are $0.''19 \times 0.''18$ and 4.39 $\mu$Jy/beam, respectively. This factor of 3 improved resolution for the east-west direction is critical to resolve the Cape of Fig. \ref{fig:B1407-425} and explore the potential slope of the spectral index toward the normal direction of the cold front. With such a high spatial resolution, we will be able to explore the spectral index distribution. We can more precisely identify the point where the anti-aging of cosmic-ray electrons happen, and will take a glimpse of where and how re-acceleration is occurring. 

Polarization observation will give the information of magnetic-field structure and hence elucidate the conditions for magnetic reconnection. Since the reconnection site would have coherent (aligned) magnetic fields and high polarization fraction, as also seen in B0600-399 and B1407-425, tangible detection of linear polarization can be made with both AA* and AA4. SKA's broadband capability, LOW, Band1, and Band2, will dramatically improve the quality of Faraday tomography \citep[see e.g.][]{2014PASJ...66...65A}, allowing us to conduct 3D view \citep{2025ApJ...992...14S} of the diffuse emission along the cold front. Those observed data is very useful to compare and validate with the data of numerical simulations of merging galaxy clusters. 

Finally, we briefly discuss whether low-$\beta$ regions might arise elsewhere in galaxy clusters, beyond cold fronts. For the density and the dynamical timescale of typical galaxy clusters, radiative cooling due to X-ray thermal bremsstrahlung is inefficient and can therefore be neglected. The system is thus assumed to evolve adiabatically. For a monoatomic ideal gas, the temperature scales with number density as $T \propto n^{2/3}$. Assuming that the magnetic-field strength follows a power-law dependence on density, $B \propto n^\delta$, $\beta$ can be written as
\begin{equation}
\beta \propto nT/B^2 \propto n^{5/3 - 2\delta}.
\end{equation}
Under the ideal MHD, compression yields $\delta = 1/3$ (compression perpendicular to the field lines) and $\delta = 2/3$ (isotropic compression). Thus, the exponent, $5/3-2\delta$, remains positive and $\beta$ increases under compression. This indicates that additional mechanisms capable of amplifying the magnetic field more efficiently than density are required to achieve the low-beta condition. Representative mechanisms include velocity shear in which magnetic-field lines are stretched, leading to amplification beyond simple compressive scaling. This mechanism was addressed in this paper. Another mechanism could be turbulence-driven amplification, where small-scale dynamo action operates through stretching and folding of magnetic field lines, resulting in exponential growth of magnetic energy. Indeed, low-$\beta$ regions are predicted in the wakes of merging clusters (blue dashed circles), where turbulent motions develop (Fig. \ref{fig:MHD}). In both scenarios, low-$\beta$ plasma regions can form, but the latter is likely to be more transient. Moreover, the latter forms the filamentary low-$\beta$ structures that may not consist of current sheets, so that interactions between such filaments and/or external action such as motions of galaxies and AGN jets may be crucial for triggering magnetic reconnection.

\section{Summary}

Using MeerKAT, GMRT, and ATCA, we found diffuse radio emissions along the cold fronts in galaxy clusters. Flattening of the spectral index implies cosmic-ray re-acceleration at the cold front, at which conventional Fermi acceleration is not expected. We propose magnetic reconnection as the re-acceleration mechanism. Although magnetic reconnection unlikely happens in ICM as ideal MHD with high $R_{\rm M}$, it could happen if resistive MHD is realized and the plasmoid instability produces fast reconnection. The plasma $\beta \sim 1$ is the condition. While there is $O(1)$~$\mu$G of ambient ICM, the reconnection site would host $O(10)$~$\mu$G to satisfy the condition. PIC simulations for turbulent flows in galaxy clusters showed that magnetic reconnection can occur if the intracluster medium (ICM) has a low plasma $\beta$ value of $\sim 1$. Actually, MHD simulations of merging galaxy clusters have shown that the compression and stretching of magnetized plasma can locally produce a low plasma $\beta$ and form a current sheet near the cold front. With SKAO AA* and AA4, we will be able to constrain where and how re-acceleration is occurring. If magnetic reconnection is real, this would be a new paradigm for the non-thermal aspects of galaxy clusters. 

\bibliographystyle{abbrvnat-maxbibnames4}
\bibliography{chapter}

@ARTICLE{2010ApJ...717..908Z,
       author = {{ZuHone}, J.~A. and {Markevitch}, M. and {Johnson}, R.~E.},
        title = "{Stirring Up the Pot: Can Cooling Flows in Galaxy Clusters be Quenched by Gas Sloshing?}",
      journal = {\apj},
         year = 2010,
        month = jul,
       volume = {717},
       number = {2},
        pages = {908-928},
          doi = {10.1088/0004-637X/717/2/908},
archivePrefix = {arXiv},
       eprint = {0912.0237},
 primaryClass = {astro-ph.CO},
       adsurl = {https://ui.adsabs.harvard.edu/abs/2010ApJ...717..908Z}
}

@ARTICLE{2012A&ARv..20...54F,
       author = {{Feretti}, Luigina and {Giovannini}, Gabriele and {Govoni}, Federica and {Murgia}, Matteo},
        title = "{Clusters of galaxies: observational properties of the diffuse radio emission}",
      journal = {\aapr},
         year = 2012,
        month = may,
       volume = {20},
          eid = {54},
        pages = {54},
          doi = {10.1007/s00159-012-0054-z},
archivePrefix = {arXiv},
       eprint = {1205.1919},
 primaryClass = {astro-ph.CO},
       adsurl = {https://ui.adsabs.harvard.edu/abs/2012A&ARv..20...54F}
}

@ARTICLE{2018PASJ...70R...2A,
       author = {{Akahori}, Takuya and {Nakanishi}, Hiroyuki and {Sofue}, Yoshiaki and {Fujita}, Yutaka and {Ichiki}, Kiyotomo and {Ideguchi}, Shinsuke and {Kameya}, Osamu and {Kudoh}, Takahiro and {Kudoh}, Yuki and {Machida}, Mami and {Miyashita}, Yoshimitsu and {Ohno}, Hiroshi and {Ozawa}, Takeaki and {Takahashi}, Keitaro and {Takizawa}, Motokazu and {Yamazaki}, Dai G.},
        title = "{Cosmic magnetism in centimeter- and meter-wavelength radio astronomy}",
      journal = {\pasj},
         year = 2018,
        month = jan,
       volume = {70},
       number = {1},
          eid = {R2},
        pages = {R2},
          doi = {10.1093/pasj/psx123},
archivePrefix = {arXiv},
       eprint = {1709.02072},
 primaryClass = {astro-ph.HE},
       adsurl = {https://ui.adsabs.harvard.edu/abs/2018PASJ...70R...2A}
}

@ARTICLE{2019SSRv..215...16V,
       author = {{van Weeren}, R.~J. and {de Gasperin}, F. and {Akamatsu}, H. and {Br{\"u}ggen}, M. and {Feretti}, L. and {Kang}, H. and {Stroe}, A. and {Zandanel}, F.},
        title = "{Diffuse Radio Emission from Galaxy Clusters}",
      journal = {\ssr},
         year = 2019,
        month = feb,
       volume = {215},
       number = {1},
          eid = {16},
        pages = {16},
          doi = {10.1007/s11214-019-0584-z},
archivePrefix = {arXiv},
       eprint = {1901.04496},
 primaryClass = {astro-ph.HE},
       adsurl = {https://ui.adsabs.harvard.edu/abs/2019SSRv..215...16V}
}

@ARTICLE{2021Natur.593...47C,
       author = {{Chibueze}, James O. and {Sakemi}, Haruka and {Ohmura}, Takumi and {Machida}, Mami and {Akamatsu}, Hiroki and {Akahori}, Takuya and {Nakanishi}, Hiroyuki and {Parekh}, Viral and {van Rooyen}, Ruby and {Takeuchi}, Tsutomu T.},
        title = "{Jets from MRC 0600-399 bent by magnetic fields in the cluster Abell 3376}",
      journal = {\nat},
         year = 2021,
        month = may,
       volume = {593},
       number = {7857},
        pages = {47-50},
          doi = {10.1038/s41586-021-03434-1},
archivePrefix = {arXiv},
       eprint = {2106.13049},
 primaryClass = {astro-ph.GA},
       adsurl = {https://ui.adsabs.harvard.edu/abs/2021Natur.593...47C}
}

@ARTICLE{2025ApJ...992...14S,
       author = {{Sakemi}, Haruka and {Chibueze}, James O. and {Cotton}, William D. and {Parekh}, Viral and {Ohmura}, Takumi and {Machida}, Mami and {Igarashi}, Taichi and {Akahori}, Takuya and {Akamatsu}, Hiroki and {Nakanishi}, Hiroyuki and {Takeuchi}, Tsutomu T.},
        title = "{Three-dimensional Polarimetric Structure of Jets from Radio Galaxy MRC 0600{\textendash}399 in A3376}",
      journal = {\apj},
         year = 2025,
        month = oct,
       volume = {992},
       number = {1},
          eid = {14},
        pages = {14},
          doi = {10.3847/1538-4357/adfc64},
archivePrefix = {arXiv},
       eprint = {2508.13454},
 primaryClass = {astro-ph.HE},
       adsurl = {https://ui.adsabs.harvard.edu/abs/2025ApJ...992...14S}
}

@ARTICLE{1958AnPhy...3..347N,
       author = {{Newcomb}, William A.},
        title = "{Motion of magnetic lines of force}",
      journal = {Annals of Physics},
         year = 1958,
        month = apr,
       volume = {3},
       number = {4},
        pages = {347-385},
          doi = {10.1016/0003-4916(58)90024-1},
       adsurl = {https://ui.adsabs.harvard.edu/abs/1958AnPhy...3..347N}
}

@ARTICLE{2025ApJ...979L..15G,
       author = {{Ghosh}, Subham and {Bhat}, Pallavi},
        title = "{Magnetic Reconnection: An Alternative Explanation of Radio Emission in Galaxy Clusters}",
      journal = {\apjl},
         year = 2025,
        month = jan,
       volume = {979},
       number = {1},
          eid = {L15},
        pages = {L15},
          doi = {10.3847/2041-8213/ad9f2d},
archivePrefix = {arXiv},
       eprint = {2407.11156},
 primaryClass = {astro-ph.CO},
       adsurl = {https://ui.adsabs.harvard.edu/abs/2025ApJ...979L..15G}
}

@ARTICLE{2007ApJ...663..816A,
       author = {{Asai}, N. and {Fukuda}, N. and {Matsumoto}, R.},
        title = "{Three-dimensional Magnetohydrodynamic Simulations of Cold Fronts in Magnetically Turbulent ICM}",
      journal = {\apj},
         year = 2007,
        month = jul,
       volume = {663},
       number = {2},
        pages = {816-823},
          doi = {10.1086/518235},
archivePrefix = {arXiv},
       eprint = {astro-ph/0703536},
 primaryClass = {astro-ph},
       adsurl = {https://ui.adsabs.harvard.edu/abs/2007ApJ...663..816A}
}

@ARTICLE{2008ApJ...687..951T,
       author = {{Takizawa}, Motokazu},
        title = "{N-Body + Magnetohydrodynamical Simulations of Merging Clusters of Galaxies: Characteristic Magnetic Field Structures Generated by Bulk Flow Motion}",
      journal = {\apj},
         year = 2008,
        month = nov,
       volume = {687},
       number = {2},
        pages = {951-958},
          doi = {10.1086/592059},
archivePrefix = {arXiv},
       eprint = {0807.3765},
 primaryClass = {astro-ph},
       adsurl = {https://ui.adsabs.harvard.edu/abs/2008ApJ...687..951T}
}

@ARTICLE{2013MNRAS.430.3249M,
       author = {{Machado}, Rubens E.~G. and {Lima Neto}, Gast{\~a}o B.},
        title = "{Simulations of the merging galaxy cluster Abell 3376}",
      journal = {\mnras},
         year = 2013,
        month = apr,
       volume = {430},
       number = {4},
        pages = {3249-3260},
          doi = {10.1093/mnras/stt127},
archivePrefix = {arXiv},
       eprint = {1301.4434},
 primaryClass = {astro-ph.CO},
       adsurl = {https://ui.adsabs.harvard.edu/abs/2013MNRAS.430.3249M}
}

@ARTICLE{2023PASJ...75S..97C,
       author = {{Chibueze}, James O. and {Akamatsu}, Hiroki and {Parekh}, Viral and {Sakemi}, Haruka and {Ohmura}, Takumi and {van Rooyen}, Ruby and {Akahori}, Takuya and {Nakanishi}, Hiroyuki and {Machida}, Mami and {Takeuchi}, Tsutomu T. and {Smirnov}, Oleg and {Kleiner}, Dane and {Maccagni}, Filippo M.},
        title = "{MeerKAT's view of double radio relic galaxy cluster Abell 3376}",
      journal = {\pasj},
         year = 2023,
        month = feb,
       volume = {75},
        pages = {S97-S107},
          doi = {10.1093/pasj/psac009},
       adsurl = {https://ui.adsabs.harvard.edu/abs/2023PASJ...75S..97C}
}

@ARTICLE{2014PASJ...66...65A,
       author = {{Akahori}, Takuya and {Kumazaki}, Kohei and {Takahashi}, Keitaro and {Ryu}, Dongsu},
        title = "{Exploring the intergalactic magnetic field by means of Faraday tomography}",
      journal = {\pasj},
         year = 2014,
        month = jun,
       volume = {66},
       number = {3},
          eid = {65},
        pages = {65},
          doi = {10.1093/pasj/psu033},
archivePrefix = {arXiv},
       eprint = {1403.0325},
 primaryClass = {astro-ph.CO},
       adsurl = {https://ui.adsabs.harvard.edu/abs/2014PASJ...66...65A}
}

@ARTICLE{2001ApJ...550.1119N,
       author = {{Nitta}, S. and {Tanuma}, S. and {Shibata}, K. and {Maezawa}, K.},
        title = "{Fast Magnetic Reconnection in Free Space: Self-similar Evolution Process}",
      journal = {\apj},
         year = 2001,
        month = apr,
       volume = {550},
       number = {2},
        pages = {1119-1130},
          doi = {10.1086/319774},
archivePrefix = {arXiv},
       eprint = {astro-ph/0011343},
 primaryClass = {astro-ph},
       adsurl = {https://ui.adsabs.harvard.edu/abs/2001ApJ...550.1119N}
}

@ARTICLE{2007ApJ...663..610N,
       author = {{Nitta}, Shin-ya},
        title = "{Continuous Transition from Fast Magnetic Reconnection to Slow Reconnection and Change of the Reconnection System Structure}",
      journal = {\apj},
         year = 2007,
        month = jul,
       volume = {663},
       number = {1},
        pages = {610-624},
          doi = {10.1086/518104},
archivePrefix = {arXiv},
       eprint = {astro-ph/0703008},
 primaryClass = {astro-ph},
       adsurl = {https://ui.adsabs.harvard.edu/abs/2007ApJ...663..610N}
}

@ARTICLE{2023PASJ...75S.138K,
       author = {{Kurahara}, Kohei and {Akahori}, Takuya and {Kale}, Ruta and {Akamatsu}, Hiroki and {Fujita}, Yutaka and {Gu}, Liyi and {Intema}, Huib and {Nakazawa}, Kazuhiro and {Okabe}, Nobuhiro and {Omiya}, Yuki and {Parekh}, Viral and {Shimwell}, Timothy and {Takizawa}, Motokazu and {Van Weeren}, Reinout J.},
        title = "{Diffuse radio source candidate in CIZA J1358.9-4750}",
      journal = {\pasj},
         year = 2023,
        month = feb,
       volume = {75},
        pages = {S138-S153},
          doi = {10.1093/pasj/psac098},
archivePrefix = {arXiv},
       eprint = {2208.04750},
 primaryClass = {astro-ph.HE},
       adsurl = {https://ui.adsabs.harvard.edu/abs/2023PASJ...75S.138K}
}

@ARTICLE{2024PASJ...76L...8K,
       author = {{Kurahara}, Kohei and {Akahori}, Takuya and {Oki}, Aika and {Omiya}, Yuki and {Nakazawa}, Kazuhiro},
        title = "{Discovery of diffuse radio source in Abell 1060}",
      journal = {\pasj},
         year = 2024,
        month = apr,
       volume = {76},
       number = {2},
        pages = {L8-L13},
          doi = {10.1093/pasj/psae011},
archivePrefix = {arXiv},
       eprint = {2311.08693},
 primaryClass = {astro-ph.HE},
       adsurl = {https://ui.adsabs.harvard.edu/abs/2024PASJ...76L...8K}
}

@ARTICLE{2026PASJ...78..137K,
       author = {{Kurahara}, Kohei and {Akahori}, Takuya and {Ohmura}, Takumi and {Yoshiura}, Shintaro and {Ito}, Daisuke and {Ma}, Yik ki and {Nakazawa}, Kazuhiro and {Omiya}, Yuki and {Sakai}, Kosei and {Sakemi}, Haruka and {Takizawa}, Motokazu},
        title = "{Origin and evolution of the {\ensuremath{\Omega}} structure in the head-tail radio galaxy of Abell 3322}",
      journal = {\pasj},
         year = 2026,
        month = feb,
       volume = {78},
       number = {1},
        pages = {137-150},
          doi = {10.1093/pasj/psaf128},
archivePrefix = {arXiv},
       eprint = {2511.03937},
 primaryClass = {astro-ph.HE},
       adsurl = {https://ui.adsabs.harvard.edu/abs/2026PASJ...78..137K}
}

@incollection{Vernstrom01.2026.SKA, author = {Tessa Vernstrom and Jennifer L West  and Cathy Horellou},title = {},year = {2026},publisher = {},note = {arXiv search: Report number AASKAII/Vernstrom01},booktitle = {Advancing Astrophysics with the SKA -- II (AASKAII)}}

@incollection{Vacca01.2026.SKA, author = {Valentina Vacca and author2 and author3 and author4 and author5},title = {},year = {2026},publisher = {},note = {arXiv search: Report number AASKAII/Vacca01},booktitle = {Advancing Astrophysics with the SKA -- II (AASKAII)}}

@incollection{AritraBasu01.2026.SKA, author = {Aritra Basu and author2 and author3 and author4 and author5},title = {},year = {2026},publisher = {},note = {arXiv search: Report number AASKAII/AritraBasu01},booktitle = {Advancing Astrophysics with the SKA -- II (AASKAII)}}

@incollection{Kurahara01.2026.SKA, author = {Kohei Kurahara and author2 and author3 and author4 and author5},title = {},year = {2026},publisher = {},note = {arXiv search: Report number AASKAII/Kurahara01},booktitle = {Advancing Astrophysics with the SKA -- II (AASKAII)}}

\end{document}